\documentclass[twocolumn]{aa}
\usepackage{graphicx}
\usepackage{psfig}
\newcommand{\gapr}{\raisebox{-.6ex}{\mbox{
$\stackrel{>}{\mbox{\scriptsize$\sim$}}\:$}}}
\newcommand{\lapr}{\raisebox{-.6ex}{\mbox{
$\stackrel{<}{\mbox{\scriptsize$\sim$}}\:$}}}
\newcommand{\modi}[1]{{#1}}

\begin{document}
\title{The thermal radiation of the isolated neutron star 
       RX\,J1856.5--3754 observed with Chandra and XMM-Newton}
\author{ V. Burwitz
\and F. Haberl
\and R. Neuh\"auser
\and P. Predehl
\and J. Tr\"umper
\and V.~E. Zavlin
 }
\offprints{Vadim Burwitz, \hspace*{\fill}\\
\email{burwitz@mpe.mpg.de}}

\institute{Max-Planck-Institut f\"ur extraterrestrische Physik, 
P.O. Box 1312, D-85741 Garching, Germany}

\date{Received October 24, 2002; accepted November 25, 2002}

\titlerunning{The thermal radiation of  
              RX\,J1856.5--3754 observed with Chandra and XMM}
\authorrunning{V. Burwitz, et al.}

\abstract{
We present results of the analysis of  data collected in 57-ks {\em XMM-Newton} 
and 505-ks {\em Chandra} observations of the nearby ($\simeq$\,120\,pc) isolated 
neutron star RX\,J1856.5--3754. We confirm most of the statements made by
Burwitz et al. (\cite{Buretal01}) who discussed the original 55-ks {\em Chandra} 
data. Detailed spectral analysis of the combined X-ray and optical data rules 
out the currently available nonmagnetic light and heavy element neutron star 
atmosphere (LTE) models with hydrogen, helium, iron and solar compositions.
We find that strongly magnetized atmosphere models also are unable to represent 
the data. The X-ray and optical data show no spectral features and are best 
fitted with a two-component blackbody model with 
$kT_{\rm bb,X}^{\infty}$\,$\simeq $\,63.5\,eV and 
 $R_{\rm bb,X}^{\infty}$\,$\simeq $\,4.4\,($d$/120\,pc) km
for the hot X-ray emitting region, and 
$kT_{\rm bb,opt}^{\infty}$\,$<$\,33\,eV and 
 $R_{\rm bb,opt}^{\infty}$\,$>$\,17\,($d$/120\,pc)\,km for 
the rest of the neutron star surface responsible for the optical flux.
The large number of counts collected with {\em XMM-Newton} allows us to 
reduce the upper limit on periodic variation in the X-ray range down to 
1.3\% (at a 2$\sigma $ confidence level) in the $10^{-3}-50$\,Hz frequency 
range. In an attempt to explain this small variability, we discuss an 
one-component model with 
$kT_{\rm bb}^{\infty}$\,$\simeq $\,63\,eV and
 $R_{\rm bb}^{\infty}$\,$\simeq $\,12.3\,($d$/120\,pc)\,km. 
This model requires a low radiative efficiency in the X-ray domain, which may 
be expected if the neutron star has a condensed matter surface.
}
  \maketitle
%
 
\section{Introduction}

\subsection{Thermal radiation from isolated neutron stars}

Thermal emission from the surface of an isolated neutron 
star (NS) could be very useful  for determining the star's 
mass $M$  and the radius $R$, what in turn puts important 
contraints on the equation of state of matter at supranuclear 
densities. 

Specifically, the detection of spectral features in radiation
of isolated NSs may provide:
\begin{itemize}
\item  
chemical composition of the NS surface,
\item  the surface gravity from the line broadening,
\item  
the mass-to-radius ratio from measurement of 
gravitational redshift of the lines,
\item
strength of the surface magnetic field.
\end{itemize}

{\it ROSAT}  was the first satellite with a sufficient sensitivity
in the soft X-ray band ($0.1-2.0$\,keV) to start a systematic 
search and study of isolated NSs which reveal blackbody-like 
thermal emission.
Seven such objects emitting soft thermal X-ray spectra, having
high X-ray/optical flux and invisible in radio band
were detected with {\em ROSAT} (see, e.~g., Zampieri et al. \cite{Zametal01} 
for the complete list).
RX\,J1856.5--3754 (or RXJ1856 throughout the text) is the brightest 
of them in X-rays.

\modi{
We use apparent temperatures $T^{\infty}_{\rm bb}$ and radii
$R^{\infty}_{\rm bb}$ which are measured by a distant observer,
throughout this paper. 
The true parameters as measured at the neutron star surface are given 
by   $T_{\rm bb}$\,=\,$T^{\infty}_{\rm bb}$\,$(1-r_g/R)^{-1/2}$
and  $R_{\rm bb}$\,=\,$R^{\infty}_{\rm bb}$\,$(1-r_g/R)^{1/2}$
where $r_g$\,=\,$2GM/c^2$ is the Schwarzschild radius of the 
NS.
}

\subsection{Previous Observations}

\begin{table*}[t]
\caption{Journal of {\it Chandra} and {\em XMM-Newton} observations}
\label{jourobs}
\begin{tabular}{c@{\extracolsep{1.5mm}}l@{\extracolsep{0.5mm}}l@{\extracolsep{0mm}}c@{\extracolsep{1mm}}c@{\extracolsep{0mm}}r@{\extracolsep{1.0mm}}c@{\extracolsep{1mm}}l}
\hline
\hline
Obs. Id. & \multicolumn{2}{c}{Instrument (mode)} & Obs. Start & Obs. End & Exposure  & \multicolumn{1}{c}{Obs.} 
&\multicolumn{1}{c}{Remark}\\
\# & & & [UT] & [UT] & \multicolumn{1}{c}{[ksec]}  &  \multicolumn{1}{c}{Type} & \\
\hline
\multicolumn{7}{l}{\it Chandra}\\
\hline
113  & \multicolumn{2}{l}{LETGS} & 2000/03/10  &  2000/03/10  &   55.48 & GTO & \\
3382 & \multicolumn{2}{l}{LETGS} & 2001/10/08  &  2001/10/09  &  101.96 & DDT & \\
3380 & \multicolumn{2}{l}{LETGS} & 2001/10/10  &  2001/10/12  &  167.50 & DDT & \\
3381 & \multicolumn{2}{l}{LETGS} & 2001/10/12  &  2001/10/14  &  171.13 & DDT & \\
3399 & \multicolumn{2}{l}{LETGS} & 2001/10/15  &  2001/10/15  &    9.32 & DDT & \\
\hline
     &  	  &		     & {\bf total exposure}&  {\bf 505.39} &     & \\
\multicolumn{7}{l}{\it XMM-Newton}\\
\hline
0106260101  & EPIC-pn  &(SW) & 2002/04/08  &  2002/04/09  &  57.193 & GTO  & effective exposure$\sim$71\%\,=\,40\,ks\\
0106260101  & EPIC-MOS1&(TI) & 2002/04/08  &  2002/04/09  &  57.740 & GTO  & no response available yet\\
0106260101  & EPIC-MOS2&(SW) & 2002/04/08  &  2002/04/09  &  57.996 & GTO  & \\
0106260101  & RGS1     &     & 2002/04/08  &  2002/04/09  &  58.546 & GTO  & \\
0106260101  & RGS2     &     & 2002/04/08  &  2002/04/09  &  58.546 & GTO  & \\
\hline
\end{tabular}
\end{table*}
Walter et al. (\cite{Waletal96}) and Neuh\"auser et al.~(\cite{Neuetal97}) 
first concluded that RXJ1856 is an isolated NS.
The {\em ROSAT} data showed that RXJ1856  has a non-variable X-ray flux 
of $\simeq 1.5\times 10^{-11}$~erg~s$^{-1}$~cm$^{-2}$ and a soft spectrum 
with a blackbody temperature of $kT_{\rm bb}^\infty\simeq 57$\,eV. 
Very interestingly, the \modi{Planckian} shape of the spectrum was confirmed later
with the first high-resolution {\em Chandra} LETGS data 
(Burwitz et al. \cite{Buretal01}; hereafter B01).
Actually, the grating spectrum did not show any significant
deviations from a \modi{Planckian}, which are expected  in spectra 
of NS atmospheres containing heavy chemical elements.

The faint optical counterpart with $V$\,$\simeq$\,26\,mag of RXJ1856
was discovered by Walter \& Matthews~(\cite{WalMat97}) and
confirmed by Neuh\"auser et al.~(\cite{Neuetal98}).
Walter~(\cite{Wal01}) and Neuh\"auser~(\cite{Neu01}) used
the faintness of the optical counterpart and its large proper motion
of about $0.33$\,mas\,yr$^{-1}$ as further arguments
that RXJ1856 is indeed an isolated NS.
With data from three {\em HST} observations of RXJ1856,
Walter (\cite{Wal01})
determined the distance to the source
to be $d=61^{+9}_{-8}$\,pc.
A re-analysis of the same data led to a revised distance to 
$d=142^{+58}_{-39}$\,pc (Kaplan et al. \cite{KapKerAnd02}).
One more {\em HST} observation allowed Walter \& Lattimer 
(\cite{WalLat02}) to obtain an improved estimate on the 
distance, $d=117\pm 12$\,pc.

None of the previous analysis of the X-ray data yielded any 
significant periodicity. 
With the {\sl ROSAT} data Pons et al. (\cite{Ponetal02}) derived
an upper limit of 6\% on the rotational modulation of the flux
for periods in the range 0.1\,s -- 20\,s.
B01, using the data from the first 55-ks {\em Chandra} observation of RXJ1856,
put a limit of 8\% in the period
range 25\,ms -- 10$^3$\,s. 
Analysing the 505-ks {\em Chandra} data Ransom et
al. (\cite{RanGaeSla02}) and Drake et al. (\cite{Draetal02}) 
obtained 
upper limits of 4.5\% (10\,ms -- 10$^3$\,s) and  2.7\% (10\,ms -- 10$^4$\,s),
respectively.  
%
%
\begin{figure*}
   \centerline{\psfig{file=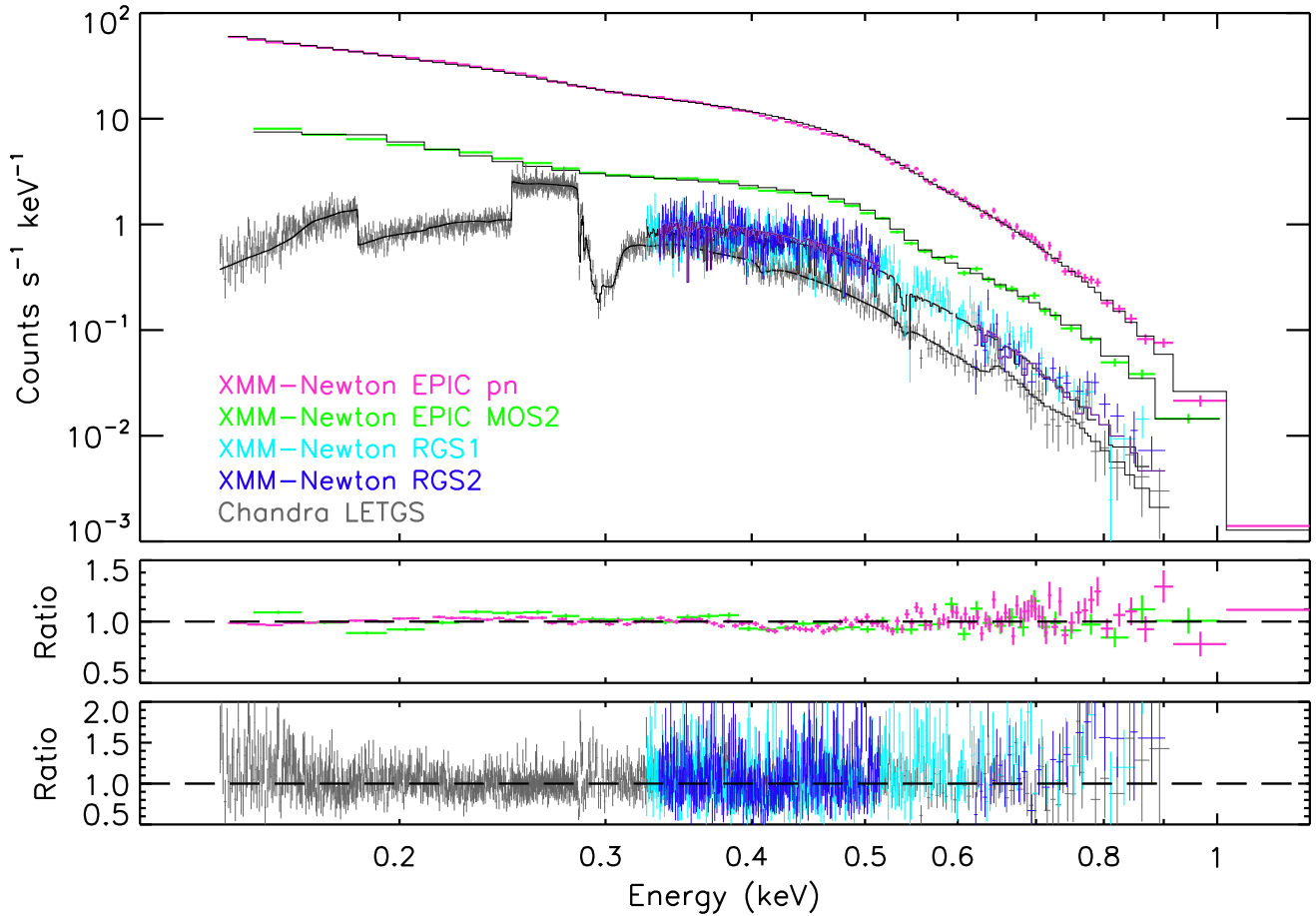,clip=,width=17.7cm} }
   \caption{\modi{The top panel shows the countrate spectra of RXJ1856
   obtained with {\sl XMM-Newton} and {\sl Chandra} with the best 
   single blackbody model fit to each instrument (the parameters 
   are given in Table~\ref{specfits}). 
   The bottom two panels show the ratio between the data and the model  
   for both the CCD detectors and the high resolution grating 
   instruments.}
\label{xmmbb} }	      
\end{figure*}
\begin{table*}
\caption{\label{specfits} Results of spectral fits to the X-ray data.}
\begin{tabular}{llr@{\extracolsep{0mm}}lr@{\extracolsep{0mm}}lr@{\extracolsep{0mm}}lccr@{/\extracolsep{0mm}}l}
\hline
Mission                   &Instrument
                          &\multicolumn{2}{c}{$n_H$}&\multicolumn{2}{c}{$kT_{\rm bb}^{\infty}$}               
			  &\multicolumn{2}{c}{$R_{\rm bb}^{\infty}$}
			  &$f_X(\rm{0.1-1.0\,keV})$   &$L_{\rm{bol}}$   
			  &\multicolumn{2}{c}{$\chi^2_{\nu}$ / d.o.f. }\\
                          &
                          &\multicolumn{2}{c}{$10^{20}$ cm$^{-2}$}&\multicolumn{2}{c}{$eV$}           
                          &\multicolumn{2}{c}{km\,($d$/120\,pc)}
			  &$10^{-11}$\,erg/cm$^2$/s  &$10^{31}$\,erg/s
			  &\multicolumn{2}{c}{}\\
\hline
\hline
{\em ROSAT} &PSPC           & 1.46&$\pm$0.20 & 56.7&$\pm$1.0 &  \hspace{3mm}7.5&$\pm$0.5  & 1.45 & 7.5 & 0.9 &   16 \\
{\em Chandra} &LETGS        & 0.95&$\pm$0.03 & 63.5&$\pm$0.2 &  \hspace{3mm}4.4&$\pm$0.1  & 1.14 & 4.1 & 1.2 & 1145 \\
{\em XMM-Newton} &EPIC-pn   & 0.18&$\pm$0.03 & 62.8&$\pm$0.3 &  \hspace{3mm}4.3&$\pm$0.1  & 1.67 & 3.7 & 2.3 &  122 \\
{\em XMM-Newton} &EPIC-MOS2 & 0.67&$\pm$0.02 & 62.6&$\pm$0.4 &  \hspace{3mm}4.4&$\pm$0.1  & 1.32 & 3.8 & 6.1 &   41 \\
{\em XMM-Newton}& RGS1+RGS2 & 0.87&$\pm$0.08 & 63.4&$\pm$0.3 &  \hspace{3mm}4.0&$\pm$0.2  & 0.90 & 3.3 & 1.1 &  717 \\
\hline
\end{tabular}
\end{table*}

In this paper we describe the {\em Chandra} and {\em XMM-Newton} 
data (\S2), the results of spectral (\S3) and timing (\S4) analysis, 
discuss implications on the nature of RXJ1856 (\S5) and propose 
alternative models (\S6).

\section{Observations and data extraction}

\subsection{{\it Chandra} LETGS}		

%
%
\begin{figure*}
   \centerline{\psfig{file=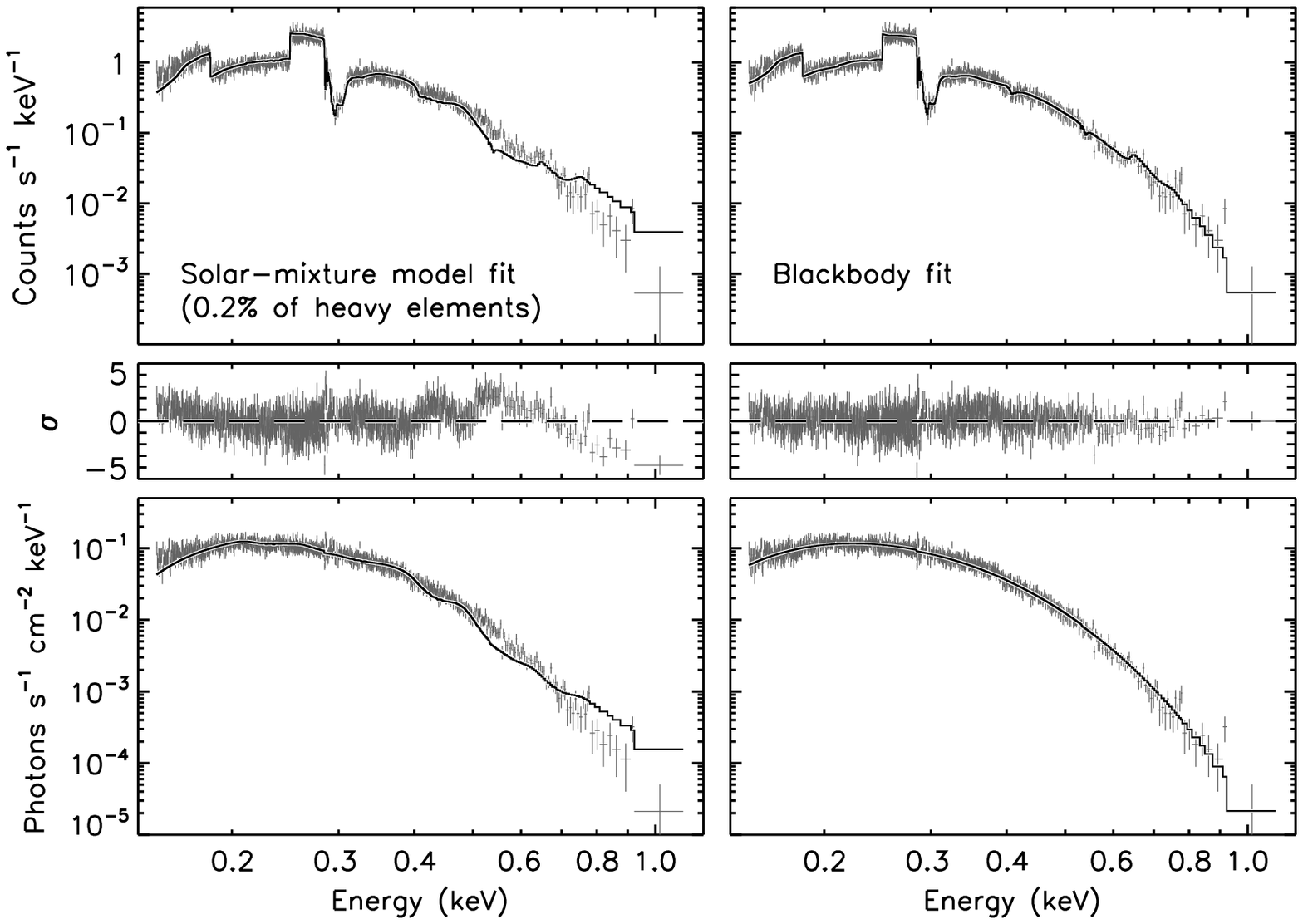,width=18.5cm,clip=} 
	       }
   \caption{\modi{The {\it Chandra}  LETGS 505\,ks X-ray spectrum fitted 
   with a NS solar-mixture atmosphere model (left) and a blackbody model (right). 
   From top to bottom the panels show the countrate spectra, the residuals given in sigmas,
    and the photon flux spectra. The model spectra are overplotted in both the top and
    bottom panels.}
   \label{atmobbmod} }
\end{figure*}
RXJ1856 was first observed in March 2000 with the standard 
LETGS configuration followed by four further observations 
in October 2001 (see Table~\ref{jourobs}).
The LETGS spectrum was extracted from the 
reprocessed level 2.0 event files. 
We used the extraction region recommended in the $Chandra$ 
Proposers' Observatory Guide\footnote{
{\tt http://asc.harvard.edu/udocs/docs/docs.html}} (POG)
for extracting the source spectrum.
For the background, large regions above and below the
source extraction area were selected (see B01 and POG)
The measured dispersed source count rate is 
0.29\,$\pm$\,0.01\,cts s$^{-1}$
in the 0.15--1.00~keV range.
The extracted source spectrum and the most up-to-date 
effective area tables (version July 2002)\footnote{{\tt 
http://asc.harvard.edu/cal/Links/Letg/User}} were used 
for spectral fits.

\subsection{XMM-Newton}

The  {\sl XMM-Newton} GTO observation took place on April 8, 2002
(see Table~\ref{jourobs}).
The duration of the exposure for all instruments was $\sim$57\,ks. 
The effective exposure for EPIC-pn CCD in the Small Window mode is 
about 70\% of this giving an effective exposure of $\sim 40$\,ks. 
The EPIC-pn and EPIC-MOS2 countrates in the 0.12-1.2 keV energy range
are $8.31\pm0.02$\,cts\,s$^{-1}$ and $1.35\pm0.01$\,cts\,s$^{-1}$, 
respectively.
Two RGS instruments detected the source at the total countrate of 
$0.29\pm0.03$\,cts\,s$^{-1}$.

\section{Spectral analysis}

B01 found that the RXJ1856 spectrum detected with the {\em Chandra} 
LETGS instument in the 55-ks observation is best fitted to a blackbody
model with  $kT_{\rm bb,X}^\infty= (63.0 \pm 1.0)$~eV and
$R_{\rm bb,X}^\infty=$\,4.4\,$\pm$\,0.5\,km\,($d$/120\,pc).
A fit to the 505-ks LETGS spectrum yields very similar,
but much more constrained values of $kT_{\rm bb,X}^\infty =63.5\pm 0.2$\,eV
and $R_{\rm bb,X}^\infty=$\,4.4\,$\pm$\,0.1\,km\,($d$/120\,pc). 
The fit is shown in Fig.~\ref{atmobbmod} (right panels).

The results of individual fits with blackbody model to the spectra 
from different instruments are given in \modi{Fig.~\ref{xmmbb} and} 
Table\,\ref{specfits}.
The {\sl Chandra} and {\sl XMM-Newton} data yield temperatures well 
consistent with each other and significantly different from that 
obtained with the {\sl ROSAT} PSPC.
The differences in the inferred values of model parameters
are attributed to systematic uncertainties in the present
calibrations of the X-ray instruments. 

\section{Timing analysis}

Details on results from the timing analysis of the {\em Chandra}
data can be found in B01, Ransom, Gaensler, \& Slane (\cite{RanGaeSla02}) 
and Drake et al. (\cite{Draetal02}).
The 57-ks observation with the {\em XMM-Newton} EPIC-pn instrument 
provided much larger statistics (370,230 counts extracted from a 
30\arcsec -radius circle centered at the source position) at a 6-ms time 
resolution.
The standard $Z^2_1$-test (Buccheri et al.~\cite{Bucetal83}) run
in the $10^{-3}-50$ Hz frequency range revealed a maximum value
$Z^2_{1,{\rm max}}=33.3$, that translates into an upper limit of
1.3\% (at a $2\sigma$ confidence level) on variability of the 
detected radiation assuming a sine-like signal.

\section{Discussion}

Pavlov et al. (1996) and Pons et al. (2002) showed that 
the light element (hydrogen and helium) nonmagnetic NS atmosphere models 
can be firmly ruled out because they, applied to 
the X-ray data on RXJ1856, yield (i) too small distance estimates ($d<10$ pc)
and (ii)
overpredict the optical flux measured from the source by a factor of 
$\sim 100$. 
As first demonstrated by B01,
no acceptable fit can be obtained with iron and
standard solar-mixture atmosphere models 
(see Zavlin \& Pavlov \cite{ZavPav02} for a recent review on 
the NS atmosphere modeling)
at any reasonable values of gravitational redshift parameter
(see B01 for a more detailed discussion). 
In case of the solar-mixture composition, the X-ray data rule
out the models with heavy element abundances greater than 0.05\%.
\modi{Fig.~\ref{atmobbmod} (left panels)} shows an example of a spectral fit with
a NS atmosphere model where heavy elements provide only
0.2\% of the total mass density. 
%
%
\begin{figure*}
   \centerline{\psfig{file=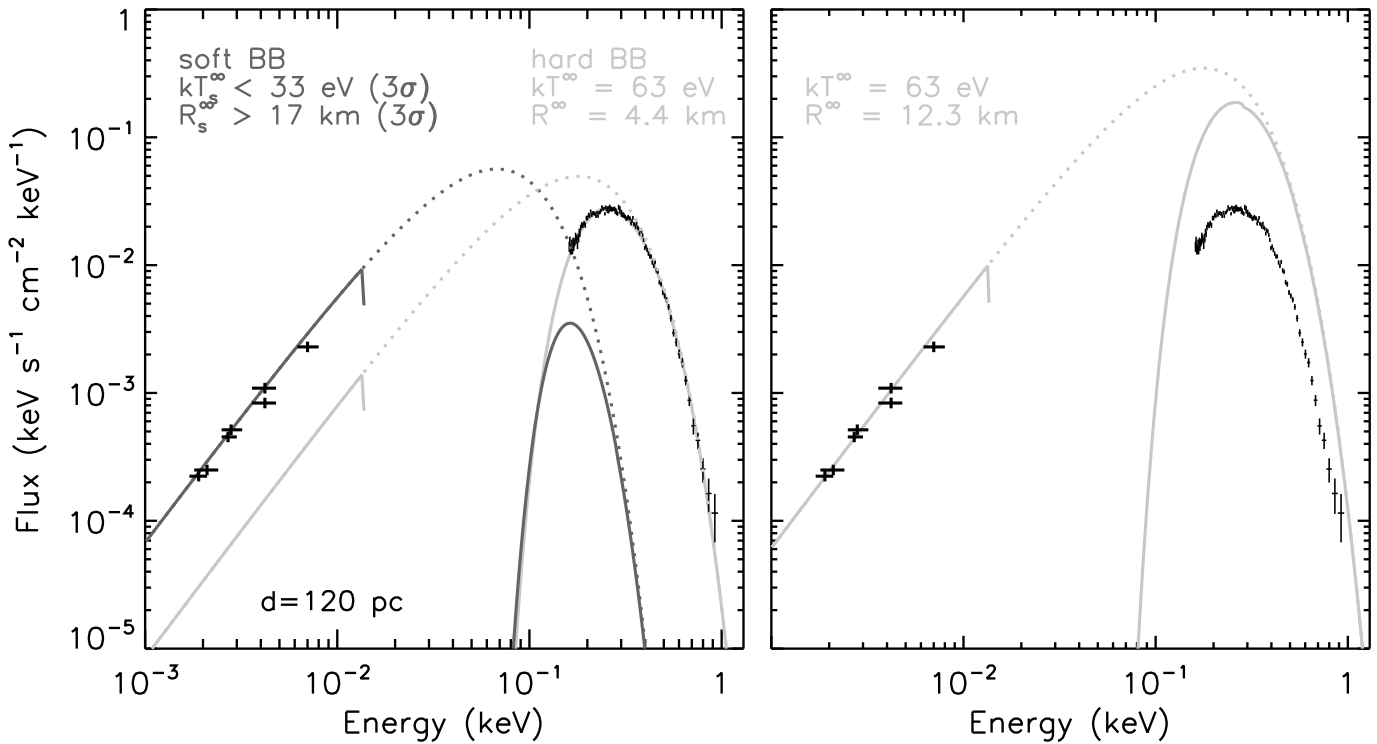,clip=,width=18.5cm}} 
   \caption{\modi{(left) Two-component blackbody model with optical 
    (van Kerkwijk \& Kulkarni \cite{KerKul01}, Pons et al. \cite{Ponetal02})
    and {\em Chandra} LETGS X-ray data.  
(right) One-component blackbody model that requires a 
    surface emissivity in the X-ray band, which is lower than that 
    of a blackbody to fit the X-ray data. 
In both panels the solid and dotted curves represent the absorbed 
and unabsorbed model spectra, respectively.}
\label{overallbb} }	      
\end{figure*}

As shown in Zavlin \& Pavlov (\cite{ZavPav02}), sharp 
features in the spectra of NS heavy-element atmospheres, 
primarily spectral lines, may be smeared out by a fast 
NS rotation at periods as short as a few milliseconds.
However, this rotational broadening (the Doppler effect)
of the spectral features yet leaves  broad-band spectral
features in the model spectra at energies around most 
prominent absorption edges, that make these atmosphere 
models fail to fit the X-ray data (Pavlov, Zavlin \& Sanwal 
\cite{PavZavSan02}; Braje \& Romani \cite{BraRom02}). 

Another possibility would be to fit the data with NS 
atmosphere models for strong magnetic fields.
However, available magnetized hydrogen models, although 
not well elaborated yet for such rather low temperatures 
of interest (see Zavlin \& Pavlov \cite{ZavPav02} for 
discussion), have the same problem as the nonmagnetic 
case: they overpredict the optical flux.
On the other hand, the spectra emitted by magnetized
iron atmospheres (Rajagopal, Romani \& Miller \cite{RajRomMil97})
show numerous absorption features which should be 
detectable with the modern instruments of high energy 
resolution.
We note that smearing of spectral features due to line 
shifts in inhomogeneous magnetic fields (for example,
varying by a factor of 2 over the NS surface for a dipole 
magnetic field configuration) may be expected to wash away  
narrow-band features, but anyway, should result in 
broad-band deviations from the blackbody spectrum, similar 
to the case of a fast rotating NS, as discussed above.

We conclude that the ``classic'' NS atmosphere models (with 
assumption of radiative equilibrium) are unable to reproduce 
the X-ray emission of RXJ1856, which is best fitted by a 
simple blackbody model (see Fig. \ref{atmobbmod}, right panels).
The possibility of a NS atmosphere which is not in radiative 
equilibrium since its outer layers heated by particle or photon
irradiation has been discussed by G\"{a}nsicke, Braje \& Romani 
(\cite{GaeBraRom02}). 
In this case one can produce a spectrum which is close to that 
of a blackbody.
However, this remains a pure speculation until the required
source of the additional heating is identified.

An alternative possibility\footnote{Originally suggested by 
G. Pavlov (see B01 for a reference).} is to assume a condensed 
matter surface --- liquid or solid ---  which might result in 
a virtually featureless \modi{Planckian} spectrum in the soft 
X-ray band.
Such a situation may occur at low temperatures ($kT$\,$<$\,86\,eV) 
and high magnetic fields ($B$\,$>$\,10$^{13}$\,G) when hydrogen,
if present on the NS surface, is expected to be in the form of 
polyatomic molecules (Lai \& Salpeter \cite{LaiSal97}; Lai \cite{Lai01}).

Yet another problem arises from the fact that the parameters 
derived from X-rays ($kT^{\infty}_{\rm bb,X}$\,=\,63\,eV 
and $R^{\infty}_{\rm bb,X}$\,=\,4.4\,($d$/120\,pc)\,km)
do not fit the optical data obeying the Rayleigh-Jeans law 
with an intensity about a factor of 7 larger than that given by the
continuation of the blackbody model yielded by the X-ray data.
This situation has led Pons et al. (\cite{Ponetal02}) to introduce
a two-component interpretation: the model applied in the X-ray band
is supplemented with an additional soft blackbody component
emitted from about 80\% of the NS surface and being responsible for 
the optical emission \modi{(c.f. Fig.~\ref{overallbb} left)}.
The requirement that the soft component does not contribute
in the X-ray band puts an upper  limit on the blackbody temperature 
$kT_{\rm bb,opt}^{\infty}$\,$<$\,33.6\,eV (a 3$\sigma$ confidence level),
that restricts the stellar radius 
$R_{\rm bb,opt}^{\infty}$\,$>$\,16.3 ($d$/120\,pc)\,km
(Pavlov, Zavlin \& Sanwal \cite{PavZavSan02}).
\modi{We also note that the spectral shape of the optical blackbody 
data puts a lower limit on the temperature of the soft component
$kT_{\rm bb,opt}^{\infty}$\,\gapr\,4\,eV 
and an upper limit on its radius
$R_{\rm bb,opt}^{\infty}$\,\lapr\,46 ($d$/120\,pc)\,km.}

The non-observed modulation of the X-ray flux imposes severe 
restrictions on the viewing geometry which has been discussed 
quantitatively by Braje \& Romani (\cite{BraRom02}). 
Using the previous limits on the pulsed fraction ($<$4\%) they 
found ''the fraction of the sky allowed by pulsed fraction 
constraints'' (which translates into a probability of a given 
orientation of the rotational and viewing axes) of 2-4\% for a 
NS radius of $\sim$\,16\,km.
The new limit established with {\em XMM-Newton} on the pulsed 
fraction reduces the allowed sky fraction to even smaller values
(around 1\%).  
Alternatively, RXJ1856 could rotate with a short period of a few
milliseconds (the available X-ray data do not provide sufficient
time resolution for seaching periodic signals at these time scales).
But this seems unlikely in view of the NS age of $\sim0.5$ Myr 
implied by its rather low surface temperature and the inferred 
distance from its birth place (Walter \& Lattimer \cite{WalLat02}).

\section{Alternative models}

The absence of periodic variations and spectral features in 
the observed radiation imposes very stringent constraints on 
any model. 
The simplest way to produce a time constant flux would be to 
assume a uniform temperature distribution across the stellar 
surface.
For a temperature of $kT^{\infty}_{\rm bb}$\,=\,63\,eV the
measured optical spectrum requires a blackbody radius 
$R_{\rm bb}^{\infty}$\,$\simeq$\,12.3\,km. 
Consequently, the X-ray emissivity has to be below that of a 
blackbody by a substantial factor.
Using the parameters derived from the {\sl Chandra} data 
(c.f. Fig.~\ref{overallbb} right) we find this factor to be about 
0.15 (it can be about 0.45 if one adopts the parameters given 
by the {\sl ROSAT} data).
This would mean that the radiating surface should have a high 
reflectivity as may be expected for a condensed matter surface
(Lenzen \& Tr\"umper \cite{LenTru78}; Brinkmann \cite{Bri80}).
In this case the spectrum may be represented by a
$\alpha_{\rm X}\,\times$\,$B_\nu$ dependence ($B_\nu$ is the 
Planck function), where $\alpha_{\rm X}$ is the absorption 
\modi{factor} ($\sim[1\,-\rho_{\rm X}]$, with $\rho_{\rm X}$ being
the reflection \modi{factor}), which in the general case will be 
energy-dependent (Brinkmann 1980).
We have tested this hypothesis by \modi{fitting the{\em Chandra} 
LETGS spectrum with a Planckian $B_\nu$ multiplied  by an energy 
dependent absorption factor $\alpha_{\rm X}$\,=\,$E^\beta$ where
$E$ is the photon energy.}
It turns out that the best fit yields $\beta=1.28\pm 0.30$, 
$kT^{\infty}_{\rm bb}=54\pm 2$\,eV, and 
\modi{$n_H$\,=\,(5.1\,$\pm$\,0.3)\, 10$^{19}$\,cm$^{-2}$.
This indicates that at a 4\,$\sigma$ level we find
find deviations from a Planckian spectrum which may result 
from an energy dependent absorption factor.}
In this case the radius required from the optical spectrum is 
$R_{\rm bb}^{\infty}\,\simeq$\,13.3\,$\alpha_{\rm opt}^{-1/2}\,
(d/120\,{\rm pc})$ km, where $\alpha_{\rm opt}$\,$\le$\,1
is the absorption \modi{factor} of the surface in the optical domain.

In conclusion, the two versions of our one-component model yield 
radii \modi{ of $R^{\infty}=12.3$\,km and 13.3\,km, respectively.
These are lower limits as the absorption factor $\alpha$ may be 
smaller than unity in the optical band.}
\modi{We note that the observed radii} correspond \modi{to true} NS radii of 
$R\,>$\,9.1\,km and $R\,>$\,10.3\,km (for a NS mass of 
$1.4\, M_\odot$), respectively, which are consistent with a soft 
equation of state. 
But, of course, the possible range of parameters allows 
stiff equations of state as well.

Two main conditions have to be fulfilled to make this model work.
Firstly, the NS has to have a condensed matter surface, which requires
a low temperature and a strong magnetic \modi{field}. The former seems to 
be fulfilled in this \modi{case: with} a temperature of $kT=54-63$ eV
RXJ1856 is the coldest one of all detected isolated NSs. But its magnetic
field is still unknown.
The second condition is that the condensed matter surface
really exhibits the required high reflectivity ($\sim 0.55-0.85$)
in the X-ray domain. It remains to be seen whether a detailed analysis of the
optical properties of magnetically condensed matter substantiates
this hypothesis.

\vskip 0.4cm
\modi{
\noindent {\em Note added in Proof}: 
Turolla, Zane \& Drake (2002) have submitted a paper to ApJ 
in which they treat the emissivity of a condensed matter
surface from a theoretical point of view, using the method of 
Brinkman (\cite{Bri80}).
}
\vskip 0.4cm

\begin{acknowledgements}

 The {\em XMM-Newton} and the {\em Chandra} LETG projects are supported by the 
 Bundesministerium f\"ur Bildung und  Forschung/Deutsches Zentrum f\"{u}r 
 Luft- und Raumfahrt (BMBF/DLR) and the Max-Planck Society.
 We would also like to thank the anonymous referee for contructive comments. 
\end{acknowledgements}

\end{document}